# Effect of Self-Lubricating Carbon Materials on the Tribological Performance of Ultra-High-Molecular-Weight Polyethylene

N. Camacho*, J.D. Escobedo-Rodríguez[1], J.M. Alvarado-Orozco, G.C. Mondragón-Rodríguez*

*CONACYT-CIDESI (CENTA and Surface Engineering and Additive Manufacturing Departments), Querétaro, Av. Pie de la Cuesta 702, 76125 Querétaro, México.
[1] Universidad Tecnológica de Querétaro, UTEQ. Av. Pie de la Cuesta 2501, 76148 Querétaro, Mexico.

**Abstract**
Ultra-high-molecular-weight polyethylene (UHMWPE) has been the gold standard for total knee replacements. Due to the knee's natural movements, UHMWPE wear debris production is inevitable. The debris, at the micron and submicron level, results in the joint's mechanical instability, reduced mobility, increased pain, and implant loosening. Wear debris production has been linked to UHMWPE mechanical properties; therefore, improvements to increase the UHMWPE mechanical properties will impact the component's longevity. Here, TiC coating and multiwalled carbon nanotubes (MWCNTs) were used to decrease the UHMWPE wear. After 400,000 cycles, the UHMWPE-MWCNT diminishes the mass loss compared to UHMWPE, and the combination with TiC decreased the material loss by ~ 43.7 % compared to the reference pair. Cold-flow and burnishing were the predominant wear modes.

*corresponding authors: guillermo.mondragon@cidesi.edu.mx

## 1. Introduction

According to the World Health Organization, ~ 23 million people endure rheumatoid arthritis, and this number will double by 2030. Arthritis is the leading condition to require a total knee replacement (TKR). TKRs consist of a femoral, tibial, and patellar components and an ultra-high-molecular-weight polyethylene (UHMWPE) articulating surface, which continues to be the gold standard for TKRs. Nevertheless, aseptic loosening due to UHMWPE wear debris-related osteolysis is the primary cause of failure in modern TKRs [1], limiting the implant's longevity. Wear debris causes mechanical instability, reduced mobility, increased pain, component loosening, and failure [2]. Reinforcement of the UHMWPE with multiwalled carbon nanotubes (MWCNTs) is a feasible alternative to improve the polymer's performance due to an increase of its mechanical properties and its free radical scavenging activity [3]. Other alternatives consider the use of solid lubricants such as carbon based coatings, i.e., carbides, diamond-like carbon (DLC). For instance, promising low wear rates of UHMWPE against DLC-coated Co-Cr-Mo were reported [4]. No polymer delamination, a negligible UHMWPE-transfer layer, and reduced UHMWPE-debris were observed. Other studies point out the controversial tribological behavior of the UHMWPE against DLCs. A. Dorner-Reisel *et al.* [5] reported that the DLC thickness is pivotal to UHMWPE wear-rate. UHMWPE wear decreased by 41.7 and 33.4 % with 0.8 and 2.7 µm DLCs on femoral Co28Cr6Mo components. However, a 4.5 µm DLC increased UHMWPE wear by 60 % compared to the uncoated component. Hauert *et al*. [6] suggested that the roughness and the testing setup (load, speed, and lubricant) also play a crucial role in UHMWPE wear. Moreover, doping of the DLC with Al, Cr or Ti is crucial to reduce internal stress and guarantee the coating adhesion. M. Jelínek et al [7] proposed that 3.3 at. % Ti might be enough to achieve proper adhesion of Ti-DLCs. The question would be how much Ti-content is enough to blend the DLC or to induce the formation of lubricant Ti carbides, since TiC has also been proved to be highly lubricant and to reduce the polyethylene wear [8].

This work aimed to investigate the potential of self-lubricating carbon-based materials to diminish the wear of UHMWPE. For this purpose, the wear behavior of UHMWPE-MWCNT nanocomposite vs. lubricant TiC was investigated. The polymer wear was analyzed under reciprocating pin-on-disk experiments up to 400,000 cycles.

2. Materials and Methods

*2.1 The synthesis of UHMWPE-MWCNTs*

Multiwall carbon nanotubes (MWCNTs) from SkySpring Nanomaterials, USA displayed a 20-30 nm Øout, 5-10 nm Øins, 10–30 µm length, and 95 % purity. Medical-grade GUR® 120 UHMWPE was donated by Ticona Engineering Polymers, Inc. Nanocomposite preparation is described elsewhere [3].

*2.2 The deposition of TiC*

The TiC coating was deposited by magnetron sputtering on mirror-polished AISI M2 steel 1" Ø × 5 mm and 6 mm 100Cr steel balls with an Oerlikon PVD coater and a 99.5 % Ti target from Plansee. First, the process started with a 2-h heating-assisted vacuum step to reach ~ 450 °C and $10^{-5}$ mbar. Then, the samples' surface was activated with Ar+ at $10^{-2}$ mbar and -50 V DC bias, 20 kHz for 40 min. The substrates were subsequently coated with a ~ 0.5 µm Ti layer by applying 7 kW of power with a constant (110 sccm) flow of high purity Ar applying 150V DC bias. Then, the ~ 0.2 µm TiN bonding coating was deposited at 7 kW and 200 sccm of high purity nitrogen at 100 V DC bias, and a top layer of ~ 0.5 µm TiC was deposited at ~ 450 °C by reactive magnetron sputtering with a mixture or pure acetylene and high purity Ar (80 : 60 sccm).

*2.3 Coating analysis*

The TiC coating was analyzed in a SmartLab XRD from Rigaku in Parallel Beam/Parallel Slit mode with ω = 2.5°, and 2θ geometry using Cu Kα radiation at 0.02° step size. The microstructure was evaluated in a Field Emission Electron Microscope (FE-SEM) JSM 7200F from JEOL.

*2.4 Surface evaluation and wear tests of UHMWPE-MWCNT vs. TiC*

All UHMWPE-MWCNT polished 10 × 10 mm specimens were weighed before testing. The roughness of the UHMWPE-MWCNTs samples was measured in a DektakXT Surface Profiler from BRUKER. For this purpose, a stylus type of 12.5 µm radius, 3 mg force, 1.5 mm distance, and 0.2 µm/pt resolution was applied. These measurements were repeated 10 times to report the average values of the polished surfaces. For comparison, the average roughnesses (Ra) of an UHMWPE articulating component manufactured by conventional machining and a mirror-polished CoCr alloy, are presented. The tribological analysis was performed in a pin-on-disk tribometer from Anton Paar. Parameters of the rotary reciprocating tests were: length stroke = 3.96 mm, 2 Hz, angle = 90°, 1N load and 100,000, 200,000 and 400,000 wear cycles. Only dry-sliding conditions were considered.

3. Results and discussion

*3.1. The surface quality of the tribological pair*

The surface roughness is critical for orthopedic applications. Dearnaley *et al.* [9] concluded that Diamond Like Coatings must be deposited onto highly polished surfaces for soft UHMWPE's optimum tribological performance. Lappalainen *et al.* [10] produced ultra-smooth coatings, reducing UHMWPE wear by factors of 40 and 600 compared to the uncoated metal. At relatively high roughnesses (few hundred nanometers), adhesion and shear strength produced at the articulating interface significantly contribute to UHMWPE wear. Here, moderate surface roughnesses were obtained. Fig. 1a shows the average roughnesses of mirror-polished UHMWPE and UHMWPE-MWCNT. The Ra of a commercial femoral and machined UHMWPE components from a recognized manufacturer were measured. The Ra of the steel balls was improved and maintained on the nm scale (average 129 nm) after TiC deposition. The SEM evaluation validated the smooth coating surface. A slightly coarser surface microstructure was observed on the coated steel balls, Fig. 1c, compared to the TiC deposited onto the flat substrates with Ra = 56 nm (Fig. 1b).

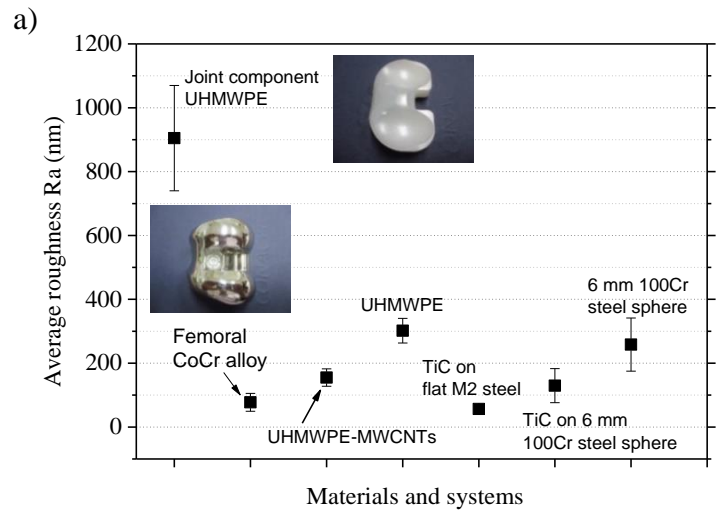

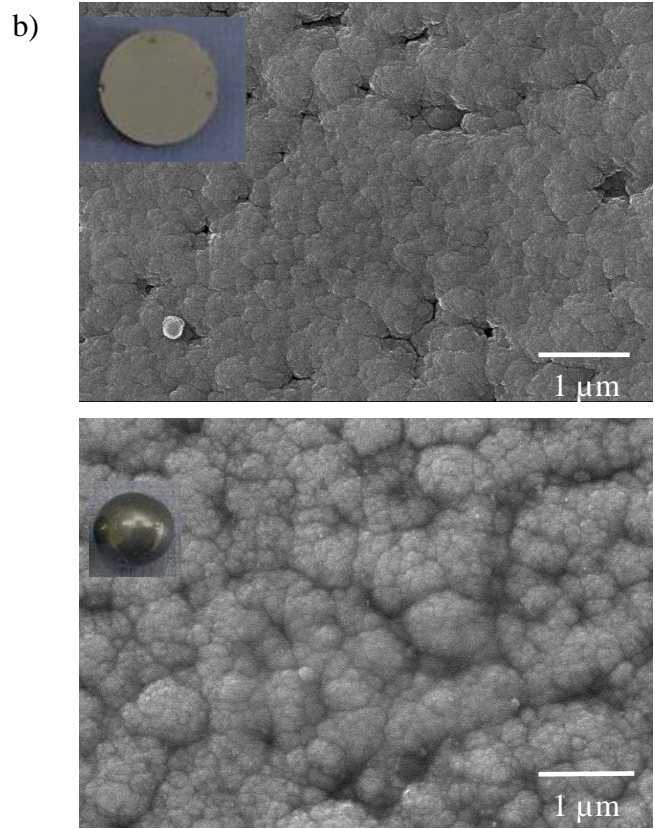

Figure 1. (a) Average roughness of UHMWPE, UHMWPE-MWCNTs, TiC on flat M2 steel and 6 mm-steel balls, UHMWPE machined, and commercial femoral components. Microstructure of the TiC on (b) flat and (c) ball specimens.

*TiC coating Microstructure and crystallinity*

The correlation of roughness, surface microstructure, chemical composition, and crystallinity of the TiC coating is crucial to explain UHMWPE-MWCNT wear behavior. Due to the deposition strategy, a high Ti-concentration was present in the top layer; this agrees with the grazing incidence XRD data shown in Fig. 2a. The XRD pattern displays c-TiC (top layer) and TiN (interlayer) peaks. Additionally, two Ti-crystalline phases (α+β), possibly related to the Ti-bottom layer, were found. Nevertheless, considering Ti's coating levels, the presence of Ti (α or β) on the top layer cannot be entirely ruled out. The coating crystallinity is a key parameter when analyzing UHMWPE wear since the ionized metal (*i.e.*, Cr, Ti) directly impacts the coating adhesion. The question that raises is how much Ti is needed to produce a

TiC?. For instance, M. Jelínek *et al.* [7] showed that 3.3 at. % Ti was needed to reach proper Ti-DLC adhesion, but nothing was mentioned regarding the formation of TiC. Additionally, Cui *et al.* [11] demonstrated that Ti remains in the metallic form in Ti-DLC coatings rather than TiC for ~ 0.41 at. % Ti, meanwhile, for Ti ~ 6.7 at. %, the formation of TiC is favored. In contrast, Wang *et al.* [12] identified TiC at a Ti-content < 1 at. %. Our EDX evaluation of the top layer in the c-TiC indicated that the Ti-content was ~ 5.1 at. %. The calculated phase diagram, using the Thermocalc methodology, of the Ti-C system indicates that mainly cubic TiC forms in the overall compositional range. For carbon content < 40 at. % the main phase is cubic TiC and only traces of hcp-Ti (or α-Ti) are present at temperatures below 500 °C. At the same temperature window and carbon contents > 40 at. % the single cubic TiC phase is thermodynamically stable. The appearance of any measurable metallic form of Ti in the top coating should be related to Ti droplets condensing in the c-TiC matrix. On the opposite, the excess of carbon may cause the formation of the graphite or even the DLC-like coating. Additional technical considerations are the carbon source and the coating chamber designs. For instance, the coatings developed by Cui *et al.* [11] and ours were deposited by magnetron sputtering in coating chambers close to 1 m$^3$ volume, but different C-source. An additional technical aspect related to the Ti-content might be the rotation strategy of the samples during coating. In our process, the c-TiC was deposited while rotating in a 2-axis planetary system.

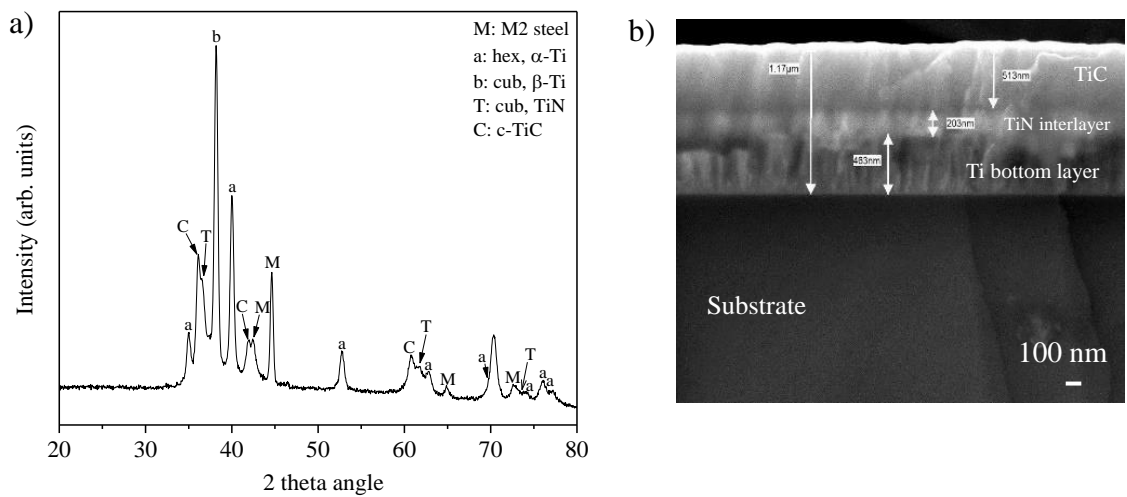

Figure 2. Grazing incidence XRD pattern (a) and cross section (b) of the c-TiC/TiN interlayer/Ti bonding coated system.

The coating thickness seems to play a critical role in polyethylene wear. A 4.7 μm DLC coating avoided the heat release and increased UHMWPE wear by 60 %, while a 0.8 μm DLC decreased the wear by ~ 41.7 % [5]. The authors reported that thin DLCs quickly dissipated heat produced by friction. Conversely, Oñate *et al.* [4] demonstrated that relatively thick DLCs (3 - 4 μm) improve UHMWPE wear; however, no heat transfer issues related to the coating thickness were mentioned. No such wear behavior of UHMWPE has been reported when bearing against TiC based coatings. The TiC thickness in this research was only ~ 0.5 μm, the whole coating system TiC/tiN/Ti ~ 1.17 μm (Fig. 2b) and no adverse wear effects on UHMWPE or UHMWPE-MWCNT were identified.

## 3.2. UHMWPE-MWCNTs vs. TiC wear

At 100,000 cycles, wear was negligible since no wear-tracks were found during the optical and the SEM evaluation. After 200,000 cycles, no mass-loss was recorded on UHMWPE and UHMWPE-MWCNT. This condition changed at 400,000 cycles. Fig. 3a shows the wear-tracks produced on UHMWPE-MWCNT vs. Ti-DLC coated counterface. Fig. 3b display pronounced wear tracks caused by the uncoated steel ball. Wear resistance improved by ~ 35 % with MWCNTs, while ~ 43.7 % was recorded for UHMWPE-MWCNT against the Ti-DLC coating. These calculations are based on the registered mass loss and wear volumes reported in Table 1. The results correlate well with the CoF of the analyzed systems. The lowest CoF (0.2) was measured for MWCNTs vs. Ti-DLC, resulting in the lowest material loss. Fig. 3a1 and 3b1 display the improvement in wear resistance, evidenced by the wear-track evaluation. A less prominent wear track in the UHMWPE-MWCNT vs. Ti-DLC compared to UHMWPE-MWCNT vs. steel system indicates the self-lubrication capacity of the MWCNTs and c-TiC. Titanium carbide is an effective solid lubricant additive [13], improving UHMWPE wear resistance [14]. Fig. 4. displays the wear-track details produced after 400,000 cycles on UHMWPE-MWCNT in the SEM-micrographs. In the UHMWPE-MWCNT vs. Ti-DLC, burnishing and scratching of the surface, and local plastic deformation in the center and along the wear track's edges were evident; only nanometer-sized debris was observed (Fig. 4a). Conversely, UHMWPE-MWCNT displays significantly more damage caused by the bare steel. There were no apparent signs of cold-flow; however, micrometer-sized debris and surface delamination indications were found.

Table 1. Weight- and volume-loss after UHMWPE-MWCNT vs c-TiC system 400,000 cycle-wear test.

| Tribological pair | Weight-loss (mg) | Volume-loss ($\mu m^3$) | Wear improvement | Wear volume/Nm |
|---|---|---|---|---|
| UHMWPE vs steel | 1.9 | 959,289.64 | Reference system | UHMWPE vs. steel: $6.0561 \times 10^{-7}$ mm$^3$/Nm |
| UHMWPE-MWCNTs vs steel | 1.2 | 623,274.35 | 35.03 % - compared to UHMWPE-Steel | MWCNT-UHMWPE vs. Steel: $3.9348 \times 10^{-7}$ mm$^3$/Nm |
| UHMWPE-MWCNTs vs c-TiC coated steel | 0.7 | 540,346.12 | 43.7 % - compared to UHMWPE-Steel 13.31 % - compared to UHMWPE-MWCNTs-Steel | MWCNT-UHMWPE vs. c-TiC: $3.41128 \times 10^{-7}$ mm$^3$/Nm |

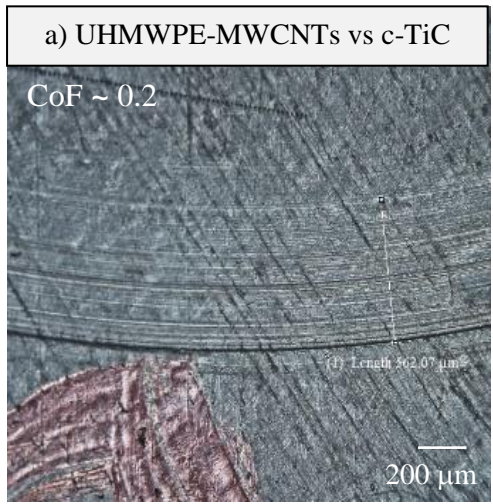 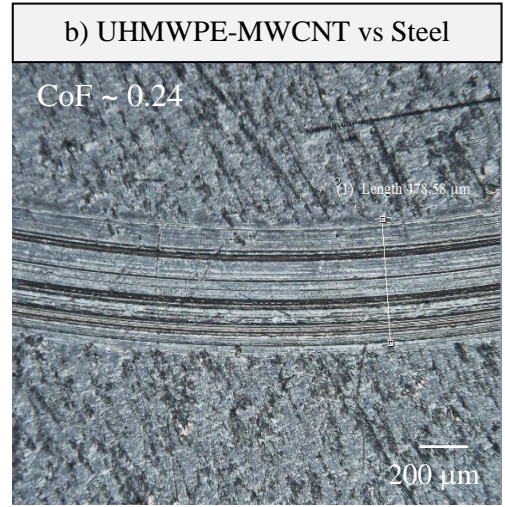

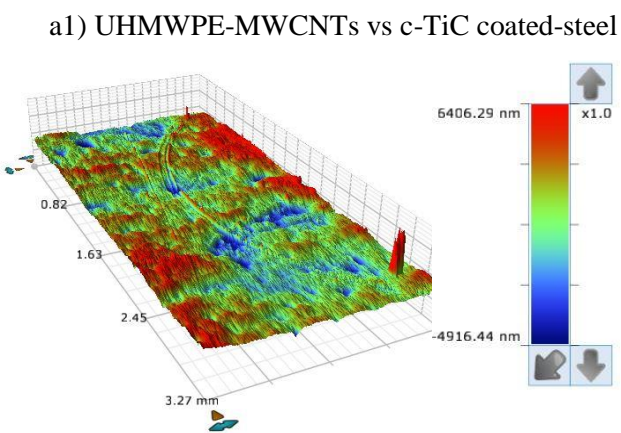 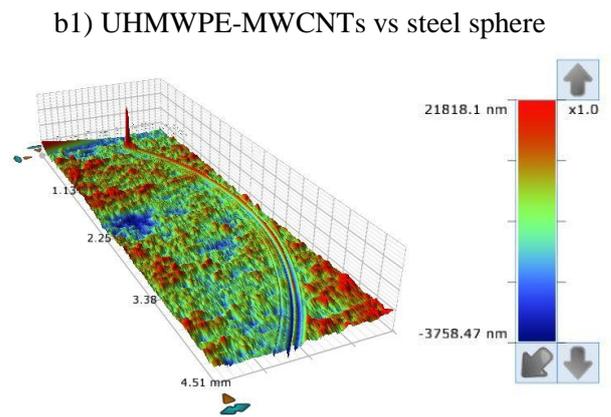

Figure 3. UHMWPE-MWCNT vs c-TiC coated steel ball (a, a1) and UHMWPE-MWCNT vs. bare steel ball (b, b1) wear tracks.

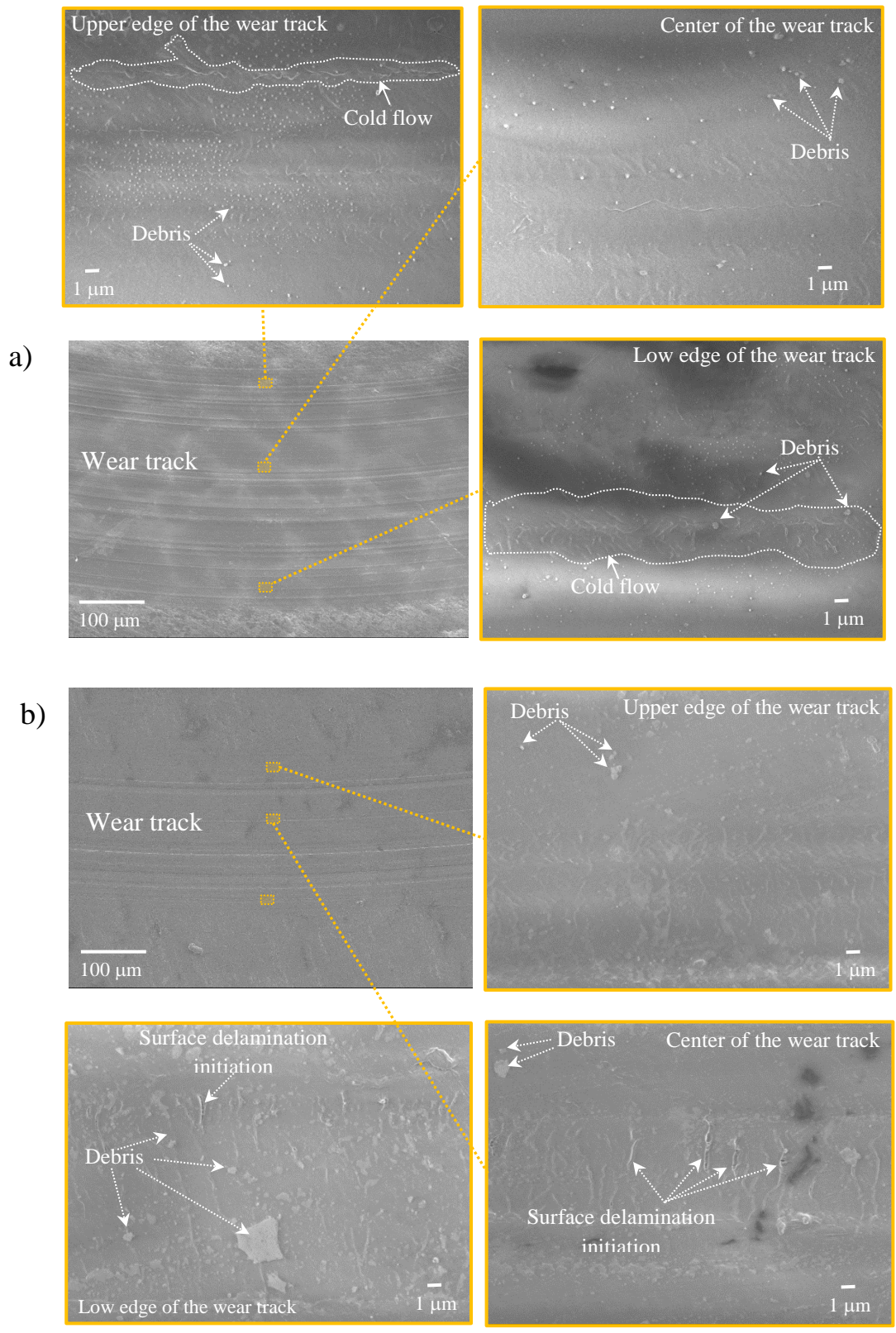

Figure 4. FE-SEM images of (a) UHMWPE-MWCNT vs c-TiC compared to (b) UHMWPE-MWCNT vs. bare steel ball wear tracks.

4. Conclusions

A reasonably smooth (~ 129 nm) self-lubricating c-TiC coating was deposited by magnetron sputtering. The coating compensated the roughness (~ 258 nm) of steel balls.

The addition of 1.25 wt. % MWCNTs improved the wear resistance of the UHMWPE by ~ 35.03 % after 400,000 cycles under dry conditions against bare steel. The homogeneous MWCNTs 3D-network around the UHMWPE powder forms a mechanical bonding that might be responsible for wear resistance improvement.

UHMWPE-MWCNT wear resistance was improved by ~ 43.7 % with the c-TiC coated steel. This behavior is mainly attributed to the coating surface finish and the lubricating nature of c-TiC.

Acknowledgments
Financial resources and materials were partially provided through the FORDECYT projects 297265 and 296384, and project 2015-02-1077.
The authors thank Ticona-Celanese Corporation for providing the UHMWPE.

Competing interest statement
The authors have no competing interest to declare.

References:
[1]   E. M. B. del Prever, A. Bistolfi, P. Bracco, L. Costa, *J. Orthop. Traumatol.*, 10 (2009), 1–8 http://doi.org/10.1007/s10195-008-0038-y.

[2]   M. Nine *et* al., *Materials*, 7 (2014) 980–1016, http://doi.org/10.3390/ma7020980.

[3]   N. Camacho, E. A. Franco-Urquiza, S. W. Stafford, *Adv. Polym. Tech.,* 37 (2018) 2261–2269, http://doi.org/10.1002/adv.21885.

 [4]   J. I. Oñate *et al.*, *Surf. Coat. Tech.*, 142–144 (2001), 1056–1062, http://doi.org/10.1016/s0257-8972(01)01074-x.

[5]   A. Dorner-Reisel, C. Schürer, E. Müller, *Diam. Relat. Mater.*, 13 (2004), 823–827, http://doi.org/10.1016/j.diamond.2003.11.080

[6]   R. Hauert, K. Thorwarth, G. Thorwarth, *Surf. Coat. Tech.*, 233 (2013), 119–130, http://doi.org/10.1016/j.surfcoat.2013.04.015.

[7]   F. Committee and F04 Committee, Test Method for Wear Testing of Polymeric Materials Used in Total Joint Prostheses. http://doi.org/10.1520/f0732-00r11.

[8]   G. Dearnaley, J. H. Arps, *Surf. Coat. Tech.*, 200 (2005), 2518–2524, 2005, http://doi.org/10.1016/j.surfcoat.2005.07.077.

[9]   R. Lappalainen, A. Anttila, H. Heinonen, *Clin. Orthop. Relat. R.*, 352 (1998), 118-127,  http://doi.org/10.1097/00003086-199807000-00014.

[10] M. Jelínek *et al.*, *Laser Phys.*, 26 (2016), 105605, http://doi.org/10.1088/1054-660x/26/10/105605.

[11] J. Cui *et* al., *App. Surf. Sci.*, 258 (2012), 5025–5030, http://doi.org/10.1016/j.apsusc.2012.01.072.

[12] Q. Wang *et al.*, *Diam. Relat. Mater.*, 25 (2012), 163–175, http://doi.org/10.1016/j.diamond.2012.03.005.

[13] Z. Liu, T. H. C. Childs, *Wear*, 193 (1996), 31–37, http://doi.org/10.1016/0043-1648(95)06652-7.

[14] Yang M. *et* al., J. Biomed. Eng., 17 (2000), 1–4.